\documentclass{article}

\usepackage{latexsym}
\newcommand{\C}{{\mathchoice {\setbox0=\hbox{$\displaystyle\rm C$}\hbox{\hbox
to0pt{\kern0.4\wd0\vrule height0.9\ht0\hss}\box0}}
{\setbox0=\hbox{$\textstyle\rm C$}\hbox{\hbox
to0pt{\kern0.4\wd0\vrule height0.9\ht0\hss}\box0}}
{\setbox0=\hbox{$\scriptstyle\rm C$}\hbox{\hbox
to0pt{\kern0.4\wd0\vrule height0.9\ht0\hss}\box0}} 
{\setbox0=\hbox{$\scriptscriptstyle\rm C$}\hbox{\hbox
to0pt{\kern0.4\wd0\vrule height0.9\ht0\hss}\box0}}}}
\newcommand{\qed}{\hfill$\Box$}

\newtheorem{theorem}{Theorem}

\newtheorem{corollary}{Corollary}
\newtheorem{fact}{Fact}

\begin{document}
\date{}

\title{\Large\bf Quantum Communication Complexity}
\author{Hartmut Klauck\\FB Informatik, Johann-Wolfgang-Goethe-Universit{\"a}t\\ 60054 Frankfurt am Main, Germany}
\maketitle

\begin{abstract}
This paper surveys the field of quantum communication complexity. Some interesting recent results are collected concerning relations to classical communication, lower bound methods, one-way communication, and applications of quantum communication complexity.
\end{abstract}

\section{Introduction}
One day Alice and Bob try to find out (while talking via their mobile phones), whether there is a CD they both like. Unfortunately the two have quite large collections (but probably not everyday taste) and so this may become very expensive. More general each player has an input from $\{0,1\}^n$ and it is their goal to compute a certain Boolean function on the concatenation of their inputs (in our example whether the two sets of CDs they like are disjoint). Surely they have to communicate to do this. We assume they have agreed previously on some rule for their communications, a so-called protocol, which describes how to compute the next message from the previous messages and the input of a player. The subject of the area of communication complexity is to find out how much communication Alice and Bob will need, and which structure their communication protocol may have to be efficient. Since both Alice and Bob are infinitely smart (they just don't know the other one's input) all internal computations are for free. This allows to concentrate on the need for communication alone. In its classical form this model has been introduced by Yao \cite{Y79} and has found many applications so far (see the monographs \cite{H97} and \cite{KN97}).

While in classical communication complexity theory the players' messages consist of bits, it is natural to ask what happens if the rules of quantum mechanics are applied, since this theory gives the most successful description of physical reality available to us. In this case the players may exchange so-called qubits, i.e., vectors in a 2 dimensional Hilbert space, the quantum mechanical analogue of bits. One may think of a qubit as a physical 2-dimensional system, implemented e.g.~ by the spin of a photon. The consideration of qubits is reasonable, because it is (at least theoretically) possible according to physics to communicate qubits, and in a way all communication is implemented physically and thus follows the laws of quantum mechanics. So we have to ask if there are any drawbacks or advantages of quantum communication. Again Yao \cite{Y93} initiated the study of quantum communication complexity. We will see that Alice and Bob can get an impressive saving when using their quantum phone!

The investigation of the power of quantum computation has attracted much interest during the last decade and is presently a very active field. The main stimulations of this interest are the breakthrough results of Shor on factoring integers in polynomial time on an quantum computer \cite{S97} and Grover on searching an item in an unordered database of $n$ elements with $O(\sqrt{n})$ queries \cite{G96}. In the meanwhile many classical computing devices have been redefined as quantum mechanical computing systems leading to interesting results about the nature of quantum computing.

It is the goal of this paper to survey quantum communication complexity results and applications of these results. Another survey on quantum communication complexity has appeared last year \cite{T99}, so we try to complement that paper with a somewhat different view.

The paper is organized as follows. In the next section we give a (rough) introduction to the definitions of quantum and classical communication complexity.
The most important topic in quantum communication complexity theory is a comparison of the power of quantum protocols to the power of classical protocols. Results on the topic are reviewed in section 3. The main applications of communication complexity theory employ its lower bounds. Thus in section 4 we look at the available methods for proving lower bounds on quantum communication complexity. In section 5 we consider one-way protocols, an important form of protocols which are easier to analyze than general protocols and still have important applications. Section 6 sketches applications to automata and formula size.

\section{The Communication Model}

In this section we provide a few definitions. 

\subsection{Quantum Mechanics}
We begin with some basics from quantum mechanics. Quantum mechanics is a theory of physical reality in terms of states and transformations of states. For general information on quantum computation see \cite{Gr99} and \cite{Pr}.

The state of a qubit is a unit vector in $\C^2$. 
 Usually an orthonormal basis of $\C^2$ is written $|0\rangle$ and $|1\rangle$, so any qubit is in a state $\alpha |0\rangle+\beta |1\rangle$ with $|\alpha|^2+|\beta|^2=1$ (also called a superposition).

The state of a vector of $k$ qubits is a unit vector in the space $\C^{2^k}$, which has an orthonormal basis
$\{|x\rangle| x\in\{0,1\}^k\}$. So the state of the $k$ qubits can be written as $\sum_i \alpha_i|x_i\rangle$ with $\sum_i|\alpha_i|^2=1$. Note that states like $\frac{1}{\sqrt{2}}(|00\rangle+|11\rangle)$ are possible, so the individual qubits do not need to be independent. This state is an example of a so-called entangled state. Generally states are written in the bra-ket notation $|\phi\rangle$.

A quantum state conveys information to us only via a measurement. For this we need an observable. For a state of $k$ qubits in $\C^{2^k}$ an observable is a set of mutually orthogonal subspaces $W_1,\ldots, W_r$, such that the direct sum of these spaces is $W_1\oplus\cdots\oplus W_r=\C^{2^k}$.

Thus any vector can be expressed as a linear combination of projections onto the $W_i$. For a state $|\phi\rangle$ the probability of getting the result $i$ when measuring according to the observable is the squared euclidean length of the projection onto subspace $W_i$. Naturally this gives us  a distribution on the possible results $\{1,\ldots,r\}$. The most common observable is the set of $2^k$ spaces spanned by the basis vectors $|x\rangle$ for $x\in\{0,1\}^k$. In this case we simply get one of the basis vectors with probability equal to the squared length of the projection onto that basis vector. A more general theory of measurements is described in \cite{Pr}.

But a measurement does not leave the state unchanged. If subspace $W_i$ is selected, then the resulting state is the normalized projection on $W_i$.

Of course measurements are not the only kind of possible transformations. A quantum state evolves by the application of a unitary transformation to the state vector.

\subsection{Classical Protocols}
Next we define classical communication protocols.

Let $f:\{0,1\}^n\times\{0,1\}^n\to\{0,1\}$ be a function. In a communication
protocol players Alice and Bob receive $x$ and $y$ and compute $f(x,y)$.
The players exchange binary messages. The communication complexity of a
protocol is the worst case number of bits exchanged.

The deterministic communication complexity $D(f)$ is the complexity of an optimal protocol
for $f$.

In a randomized protocol both players have access to public random
bits. The output is required to be correct with probability $1-\epsilon$ for
some constant $\epsilon$. The randomized communication complexity of a function
$R_\epsilon(f)$ is then defined analogously to the deterministic communication complexity.

A Las Vegas protocol is a randomized protocol in which the players never err, but may give up with some probability $\epsilon$. The complexity is denoted $R_{0,\epsilon}(f)$ and $R_0(f)=R_{0,1/2}(f)$.

In a nondeterministic protocol each player may guess privately and the players accept if some guess leads to acceptance. The complexity is denoted $N(f)$.

A protocol is called one-way if Alice sends a message and Bob immediately computes the output. The complexity notations under this restriction are superscripted like $D^1$.

The communication matrix of $f$ is $M[x,y]=f(x,y)$.

\subsection{Quantum Protocols}

Now we define quantum communication protocols.

In a quantum protocol both players have a private set of qubits each initialized to $|0\rangle$. At the beginning both receive a Boolean input string. In each communication round one player applies a unitary transformation to the qubits in his possession and then sends some of the qubits to the other player (this does not change the overall state, but rather the possession of the individual qubits).
At the end of the protocol the state of some qubits belonging to one player is measured and (a part of) the result is taken as the output.

In a (bounded error) quantum protocol the correct answer must be given with
probability $1-\epsilon$ for some $1/2>\epsilon>0$. The (bounded error) quantum
complexity of a function is called $Q_\epsilon(f)$.

In a Las Vegas quantum protocol the protocol may never err, but may give up with probability $\epsilon$. The Las Vegas quantum complexity of a function is called $Q_{0,\epsilon}(f)$ resp.~$Q_0(f)=
Q_{0,1/2}(f)$. $Q_E(f)=Q_{0,0}(f)$ is called the exact quantum complexity of a function.

In a nondeterministic quantum protocol an input is accepted iff the protocol computes output 1 with probability larger than zero, $NQ(f)$ is the nondeterministic quantum complexity.

The above model can simulate most models with stronger definitions, where e.g.~intermediate measurements are allowed. Notably different is another definition (\cite{CB97} and \cite{CDNT97}), where Alice and Bob may possess an arbitrary input-independent set of (entangled) qubits and then communicate. This model is stronger than the above: the superdense coding
technique of \cite{BW92} allows to transmit $n$ bits of classical information with
$\lceil n/2\rceil$ qubits using $n/2$ pairs of entangled qubits in state $\frac{1}{\sqrt{2}}(|00\rangle+|11\rangle)$, where the first qubit belongs to Alice and the second one to Bob. We do not discuss this model in the following, except when explicitly mentioned.
A multiparty quantum communication model has been investigated in \cite{BDHT97}.

\section{Comparing Quantum to Classical Communication}

The first question that arises is, of course, whether the quantum communication complexity of some function is smaller than the classical communication complexity. This section deals with such results. As a start we discuss a fact from quantum information theory that seems to be somewhat discouraging, Holevo's theorem \cite{H73}.

\begin{fact} Alice receives a uniformly chosen input $X$ from some finite set. If Alice and Bob communicate $t$ qubits, then for any measurement on Bob's qubits with a resulting random variable $Y$ the mutual information between $X$ and $Y$ is at most $t$.
\end{fact}

But for the same reasons why Shannon Theory does not tell us all about classical communication complexity, we are not finished yet. Alice and Bob want to know only one bit (the function value), so they do not necessarily have to exchange all their inputs.

A striking example of a speedup by quantum communication is the following result of \cite{BCW98}.
In the disjointness problem $DISJ$ Alice and Bob receive vectors $x,y\in\{0,1\}^n$. They want to know whether $\bigvee_{i=1}^n x_i\wedge y_i=1$ (actually this is the complement of testing whether the sets $x$ and $y$ are disjoint).

\begin{fact}
$Q_\epsilon(DISJ)=O(\sqrt{n}\log n)$.

$R_\epsilon(DISJ)=\Omega(n)$.
\end{fact}

The protocol is an application of Grovers quantum search algorithm \cite{G96} derived in \cite{BCW98} from a general simulation of quantum decision trees by quantum communication protocols. It takes $O(\sqrt{n})$ rounds to run the protocol. The lower bound is from \cite{KS92}.

This is the maximal gap between bounded error quantum and probabilistic communication complexity for a total function known so far.

In the Las Vegas Setting the following is known.

\begin{fact}
There is a total function $f:\{0,1\}^n\times\{0,1\}^n\to\{0,1\}$ such that

$Q_{0,\epsilon}(f)=O(n^{2/3+\delta})$ for all constants $\delta>0$.

$R_\epsilon(f)=\Omega(n/\log n)$.
\end{fact}

A polynomial gap for the function is proved in \cite{Kl00} (see also \cite{BCWZ99}), the upper bound has subsequently been improved by Ambainis \cite{A00}.

In classical communication it is known that a quadratic gap between Las Vegas and deterministic communication complexity is the maximum possible for total functions. This follows from the relation $D(f)=O(N(f)\cdot N(\neg f))=O(R^2_0(f))$, see \cite{KN97, H97}. In quantum communication such a relation is unknown. Recently de Wolf has shown that nondeterministic quantum communication can be characterized as follows: Let $M(f)$ be the communication matrix of $f$. Then the nondeterministic rank of $f$ is $nrank(f)$, the minimum rank of a matrix obtainable from $M(f)$ by replacing nonzero entries by arbitrary nonzero reals.

\begin{fact} $NQ(f)=\Theta(\log nrank(f))$.
\end{fact}

Invoking a result from \cite{L90}, which states $D(f)=O(N(\neg f)\cdot \log trank(f))$, where $trank(f)$ denotes the maximal number of rows of a rectangle in the communication matrix that contains only zeroes above a diagonal of ones, we get

\begin{theorem}
$D(f)=O(N(f)\cdot NQ(\neg f))=O(N(f)\cdot Q_0(f))$ for all total $f$.
\end{theorem}

So for superpolynomial gaps between $Q_0(f)$ and $R_0(f)$ also both $N(f)$ and $N(\neg f)$ must be large.

The most important open question of quantum communication complexity is whether there is an exponential gap between bounded error quantum communication and bounded error probabilistic communication.

For partial functions the situation is very different. The following has been proved by Raz \cite{R99}.

\begin{fact} There is a partial function $f:\{0,1\}^n\times\{0,1\}^n\to\{0,1\}^n$ such that

$Q_\epsilon(f)=O(\log n)$

$R_\epsilon(f)=\Omega(n^{1/4}/\log n)$
\end{fact}

The next result is from \cite{BCW98}.

\begin{fact} There is a partial function $f:\{0,1\}^n\times\{0,1\}^n\to\{0,1\}^n$ such that

$Q^1_E(f)=O(\log n)$.

$R_0=\Omega(n)$.

$R^1_\epsilon(f)=O(\log n)$.
\end{fact}

\section{Lower Bound Methods}

While quite a lot of lower bound methods are known for classical communication
complexity, so far only few lower bound methods for quantum protocols are known.

The following lower bound from \cite{BCW98} employs the rank of the communication matrix over the reals. For classical deterministic protocols the method has been introduced in \cite{MS82}.

\begin{fact}
$Q_E(f)=\Omega(\log rank(M(f)))$.
\end{fact}

The rank conjecture in communication complexity is that $\log rank(M(f))$ is polynomially related to the deterministic communication complexity (see \cite{KN97} for a discussion). If this is true, then exact quantum communication does not help much for total functions.

Theorem 1 gives us the lower bound $Q_0(f)=\Omega(D(f)/N(f))$ for total
functions.

The so-called discrepancy lower bound is known to hold for bounded error quantum protocols by \cite{K95}. But the latter technique is not strong enough to prove e.g.~a superlogarithmic lower bound on the bounded error quantum communication complexity of the disjointness problem. Here only a $\Omega(\log n)$ bound is known, which is nontrivial only in the model with prior entanglement (see \cite{BW99}).

Another lower bound method is developed in \cite{BW99}. This method is basically an approximate version of the rank lower bound, and is very hard to handle, only a logarithmic lower bound on disjointness is known to be provable via the method currently.

The rank lower bound and approximate rank bounds are also discussed in \cite{ASTVW98}, where nonstandard communication problems are considered. \cite{CDNT97} proves a linear lower bound on the inner product function via Holevo's theorem.

\section{One-Way Communication}

The quantum protocols discovered so far that achieve a polynomial speedup for total
functions compared to classical protocols \cite{BCW98},\cite{Kl00} share the feature of using a large amount of communication rounds (which comes from using Grover search).
We turn our interest to the question how efficient communication problems can
be solved in the quantum world when the number of rounds is restricted.
The most restricted model is one-way communication, where only a monologue is transmitted from Alice to Bob, who decides the function value.
Kremer \cite{K95} investigates quantum one-way communication and exhibits a partial function which is a complete problem for polylog quantum communication (with bounded error).

\cite{Kl00} generalizes a lower bound of \cite{KNR95} on randomized one-way communication based on the notion of the VC-dimension to the case of quantum one-way communication.
In this paper we show a new combinatorial lower bound method that is always at most a log-factor worse than the VC-dimension bound, but is sometimes much better.

For a function $f(x,y)$ a set $S$ is shattered, if for all $R\subseteq S$ there
is an $x$ such that $f(x,y)=1\iff y\in R$ for all $y\in S$.
The VC-dimension of a function $f(x,y)$ is the maximal size of a shattered set.

The following is proved in \cite{Kl00}.

\begin{fact}
For all $f: Q^1_\epsilon(f)\ge\Omega(VC(f))$.
\end{fact}

It is known that this lower bound is not always tight for probabilistic one-way communication, for the following see \cite{KNR95}, \cite{Kl00}, \cite{BW99}.

\begin{fact}
$Q_\epsilon^1(DISJ)\ge \Omega(n)$.

$R_\epsilon^1(GT)=\Omega(n)$ for the function $GT(x,y)=1\iff x\ge y$.

$VC(GT)=1$.
\end{fact}

Basically the VC-dimension is the size of the largest instance of the Index function which can be embedded in the communication matrix as shown in \cite{Kl00}.
The Index function is defined as follows: $IX_n:\{0,1\}^n\times \{0,1\}^{\log
n}\to \{0,1\}$ and $f(x,y)=1\iff x_y=1$. Note that the player possessing $x$
has to send the message in a one-way protocol.

In the following we describe a new lower bound technique for quantum one-way communication complexity.

Let $M$ be the communication matrix of $f$, $row(M)$ denotes the number of different rows in a matrix $M$. The lower bound method works as follows:

\begin{itemize}
\item Choose a subset $M'$ of the rows of $M$ so that the bound constructed in the following is as large as possible.
\item Let $h$ be minimal with the following property:

For all subsets $M''$ of the rows of $M'$ obtained by fixing $M'(r,c_i)$ for some columns $c_i$, there is a column $c$ such that:

If $U$ resp.~$V$ denotes the number of ones resp.~zeroes in $c$ over the rows in $M''$, then $\max\{U/(U+V),V/(U+V)\}\le h$.
\item Let
$bound(f)=\log row(M')/\log_2\log_{1/h} row(M')$.
\end{itemize}

\begin{theorem} For all total $f$: $Q_\epsilon^1(f)=\Omega(bound(f))$.\end{theorem}

{\sc Proof Sketch:} Assume that $P$ is a quantum one-way protocol for $f$ with error $\epsilon$ and $C$ qubits communication. The bound is attained with parameters $M',h$. First the error probability is decreased to $\epsilon/\log^2_{1/h} row(M')$ by repetition. This increases communication to $O(C\cdot \log_2\log_{1/h}row(M'))$. Now showing that $\Omega(\log_2 row(M'))$ qubits communication is needed establishes the theorem.

We argue that the protocol with the small error can be used to transmit a complete row of $M'$ to Bob with high probability (when the rows are chosen uniformly). Since there are $row(M')$ different strings to be transmitted the communication must be $\Omega(\log_2 row(M'))$ for the modified protocol and $\Omega(\log row(M')/\log_2\log_{1/h} row(M'))$ for the original protocol as a consequence of Holevo's theorem (fact 1).

Now it has to be shown that Bob, given the message can reconstruct the row given to Alice exactly with large probability. Bob proceeds as follows:

Bob tries to evaluate the message on a column. Each column is associated to an observable the protocol applies to messages to find the function value with high probability. Bob measures in order to remove as many rows as possible. This is done by taking the column that minimizes $\max\{U/(U+V),V/(U+V)\}$. This value is always at most $h$. If the evaluation is 1, then $V$ rows are removed, otherwise $U$ rows, in both cases at most a fraction of $h$ of the rows remains. So $\log_{1/h} row(M')$ questions suffice, if they are all answered correctly, to find the correct row.

An analysis as in \cite{ANTV99} shows that the probability of an error occuring in any of the measurements is at most $4\sqrt{\epsilon}$ (this is similar to the case of serial dense quantum coding, since the questions  are adaptive).\qed

First we apply the method to the Index function and yield

\begin{corollary} $bound(IX_n)=\Omega(n/\log n)$.
\end{corollary}

Now we compare the lower bound method to the VC-dimension method.

\begin{theorem}
For all total $f$: $bound(f)=\Omega(VC(f)/\log VC(f))$.
\end{theorem}

As we have seen the VC-dimension of the $GT$ function is only 1, but the new method performs better (see \cite{Kl00}).

\begin{corollary}
$Q_\epsilon^1(GT)=\Omega(n/\log n)$.
\end{corollary}

Another result of \cite{Kl00} is the following fact about quantum one-way Las Vegas protocols.

\begin{fact}
$Q_{0,\epsilon}^1(f)\ge (1-\epsilon) D^1(f)$ for all total function $f$.
\end{fact}

So interaction is crucial for a quantum/classical Las Vegas speedup, compare fact 3.

\section{Applications}

Applications of communication complexity lower bounds to quantum formula size are given in \cite{Y93}, implicitly in \cite{RV99}, and in \cite{Kl00}.

\cite{Kl00} gives lower bounds on the size of quantum finite one-way automata in terms of one-way communication complexity lower bounds (quantum automata are defined in \cite{KW97}). The results imply that quantum Las Vegas finite one-way automata are never more than polynomially smaller than dfa's (quantum bounded error finite automata are sometimes exponentially smaller and sometimes exponentially larger than dfa's for the same language).

\section{Conclusions}

Quantum communication complexity is presently a very active field. Due to the strange rules of quantum mechanics new interesting results have to be expected. The research also leads to new topics in communication complexity, such as the complexity of sampling investigated in \cite{ASTVW98}.

But so far not all classical areas of research in communication complexity have been investigated thoroughly. There is e.g.~no study of multiparty communication complexity in the blackboard model, see \cite{BNS92}. Research on the important topic of applications of quantum communication complexity lower bounds is still rare.

\end{document}